\pdfoutput=1
\documentclass[11pt]{article}
\usepackage{graphicx}

\usepackage[]{acl}

\usepackage{times}
\usepackage{latexsym}
\usepackage{natbib}
\usepackage{algorithm}
\usepackage[T1]{fontenc}
\usepackage[utf8]{inputenc}

\usepackage{microtype}
\usepackage{caption,subcaption}
\usepackage{graphicx}
\usepackage{amsmath}
\usepackage{multirow, epstopdf}
\usepackage{amssymb}
\title{\textit{BinarySelect} to Improve Accessibility of Black-Box Attack Research}

\author {Shatarupa Ghosh \and  Jonathan Rusert\thanks{~~Corresponding Author} \\Purdue University, Fort Wayne\\
    \texttt{shatarupa.ghosh012@gmail.com, jrusert@pfw.edu} \\}
\usepackage{hyperref}
\usepackage{algorithm}
\usepackage{algorithmic}
\usepackage{amsmath}
\begin{document} 
\maketitle

\vspace{1cm}

\begin{abstract}
Adversarial text attack research is useful for testing the robustness of NLP models, however, the rise of transformers has  greatly increased the time required to test attacks. Especially when researchers do not have access to adequate resources (e.g. GPUs). This can hinder attack research, as modifying one example for an attack can require hundreds of queries to a model, especially for black-box attacks. Often these attacks remove one token at a time to find the ideal one to change, requiring $n$ queries (the length of the text) right away. We propose a more efficient selection method called \textit{BinarySelect} which combines binary search and attack selection methods to greatly reduce the number of queries needed to find a token. We find that \textit{BinarySelect} only needs $\text{log}_2(n) * 2$ queries to find the first token compared to $n$ queries. We also test \textit{BinarySelect} in an attack setting against 5 classifiers across 3 datasets and find a viable tradeoff between number of queries saved and attack effectiveness. For example, on the Yelp dataset, the number of queries is reduced by 32\% (72 less) with a drop in attack effectiveness of only 5 points. We believe that \textit{BinarySelect} can help future researchers study adversarial attacks and black-box problems more efficiently and opens the door for researchers with access to less resources. 
\end{abstract}

\section{Introduction}
\begin{figure*}
    \centering
    \includegraphics[width=\textwidth]{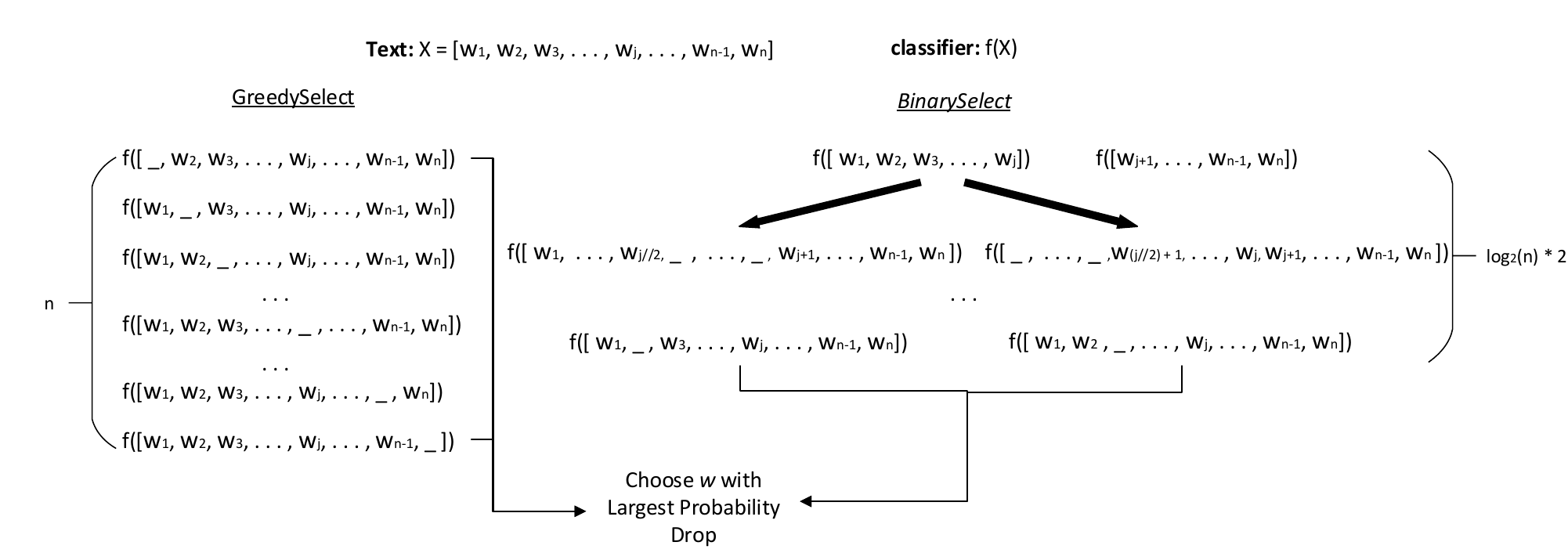}
    \caption{Visualization of GreedySelect versus \textit{BinarySelect}. GreedySelect removes 1 word at a time and checks the change in probability. \textit{BinarySelect} continuously splits the text in 2 and excludes the segments from the query. The excluded segment which causes the highest drop in target class probability is split again and so on. Eventually, the splitting leaves only 1 word which is chosen. }
    \label{fig:bsgs}
\end{figure*}
Adversarial text attacks have seen a surge in research in recent years \cite{qiu2022adversarial}. Attacks help to test the robustness of Natural Language Processing (NLP) models by both testing words and syntactic structures an NLP model might not be familiar with \cite{iyyer-etal-2018-adversarial,qi-etal-2021-mind}, as well as simulating attacks that humans may use to trick the NLP model \cite{formento-etal-2023-using,wang-etal-2022-semattack}. 

Adversarial attacks assume some level of knowledge of the models they target. White-box attacks \cite{sadrizadeh-2022-block,choi-etal-2021-outflip} have access to a model's weights and architecture. This allows the attack to more quickly find the tokens which the classifier is leveraging, however, this may be unrealistic when considering models deployed online. Black-box attacks \cite{deng-etal-2022-valcat,le-etal-2022-perturbations} have access only to the output (e.g. label) and probabilities (or logits) of a model. This restriction means that black-box attacks spend more time querying the model to find the same tokens. 

In the case of text classification, attacks often remove or mask one token (word) at a time and check the change in probability \cite{jin2020bert,li-etal-2020-bert-attack,ren-etal-2019-generating,formento-etal-2023-using}. In a text of length $n$, this results in $n$ number of queries before the attack even starts to change the text. For longer texts, this can slow down attacks. For researchers with access to fewer resources (e.g. no or few GPUs), this can greatly hinder verifying attacks, or other related research (e.g. attack defense or attack detection). In this research we propose a new method, \textit{BinarySelect}, to reduce the number of queries required to find the most relevant words\footnote{We focus on words, and verify on characters later on.} to the model.

\textit{BinarySelect} is inspired by the binary search algorithm. Whereas binary search requires a sorted list of values, \textit{BinarySelect} uses the probabilities returned from the classifier to guide its search. Specifically, \textit{BinarySelect} removes the first half of the text and compares the change in the probability to removing the second half. The half that causes the larger drop becomes the new search area and the algorithm repeats until 1 word (or token) remains. This algorithm greatly speeds up finding the first relevant word. Furthermore, to reduce future queries the algorithm leverages a binary tree to hold probabilities that have already been found. We find a tradeoff with \textit{BinarySelect} with a small reduction in effectiveness but a large reduction in number of queries. 
Though we focus on adversarial attacks, \textit{BinarySelect} could also be leveraged for other black-box NLP models.

Our research makes the following contributions:

1. Propose a new selection algorithm, \textit{BinarySelect}, to reduce the number of queries required by adversarial attacks and make attack and related research more accessible to others. 

2. Explore and verify the theoretical effectiveness of \textit{BinarySelect} in finding the most relevant words in text classification. We find that \textit{BinarySelect} is able to find the first relevant word in $\text{log}_2(n) * 2$ queries, which strongly outperforms the previous GreedySelect at $n$ queries. 

3. Evaluate \textit{BinarySelect} as a tool in adversarial attacks for 3 text classification datasets against 5 text classification systems. We find that \textit{BinarySelect} offers a strong tradeoff by reducing the average number of queries by up to 60\% with a smaller drop in attack effectiveness.

Overall \textit{BinarySelect} provides an alternative selection method for black-box algorithms. It provides an easy way to balance number of queries with attack effectiveness to allow researchers with lower resources a place in the field\footnote{Our code can be found at \url{https://github.com/JonRusert/BinarySelect}}.

%\section{Methodology}
%To establish a baseline for our experiments, we employ a sentiment analysis model pre-trained on a large corpus of text data. Specifically, we leverage the "textattack/bert-base-uncased-imdb" model, renowned for its proficiency in sentiment classification tasks. This pre-trained model is specifically tailored for sentiment analysis on the IMDB dataset. The model's proficiency stems from its meticulous fine-tuning process during its initial development. This ensures its adaptability to the specific sentiment analysis objectives we aim to achieve in this study. 

%In parallel with conventional sentiment analysis, we venture into the domain of adversarial attacks. Adversarial attacks involve the crafting of input samples with the intention of inducing misclassifications or perturbations in model predictions. We employ a suite of attack strategies, including word substitution, to exploit potential vulnerabilities in sentiment analysis models. These attacks are implemented with different levels of perturbation to assess the model's ability to withstand diverse degrees of adversarial challenges.

\section{Proposed Approach}
%In this study, we employ two distinct approaches, namely, greedy search and binary search, to elucidate the influence of specific words on sentiment analysis outcomes. 
In this section we define our proposed selection method, \textit{BinarySelect}. For background, we first define the goal of a word selection method and then define the approach commonly used by previous black-box attack research\footnote{More related work found in Appendix \ref{sect:relatedwork}}, which we call \textit{GreedySelect}. A visualization of the difference between the two methods can be found in Figure \ref{fig:bsgs}. Note that we test our proposed method in the area of text classification, so terminology focuses on this area specifically moving forward. 

\subsection{Threat Model}
The approaches assume black-box knowledge of a model. Specifically, no knowledge of model architecture or weights are known. Approaches are able to send queries to the model and the model returns a label (if classification) and confidence score. These assumptions follow previous black-box adversarial attack research in NLP \cite{alzantot2018generating, garg2020bae, li-etal-2020-bert-attack,gao2018blackbox, hsieh-etal-2019-robustness, li2021contextualized}. Note some prior research has referred to this as ``grey-box'' due to probability access.

\subsection{Selection Methods} Let a text of length $n$, be represented as $X = \{w_1, w_2, ..., w_n\}$, where $w_i$ is the $i$-th word in the text.  The goal of a selection method is to return the word $w_j$ (or token) which has the greatest impact on a classifier's decision (or probability). Note that $w_j$ is generally then replaced with a new word or modified to hurt the classifier in its ability to make the best decision. After replacement, the selection method then returns the word with the second-highest impact on the classifier's decision, and so on. 
\begin{figure*}
    \centering
    \includegraphics[width=\textwidth]{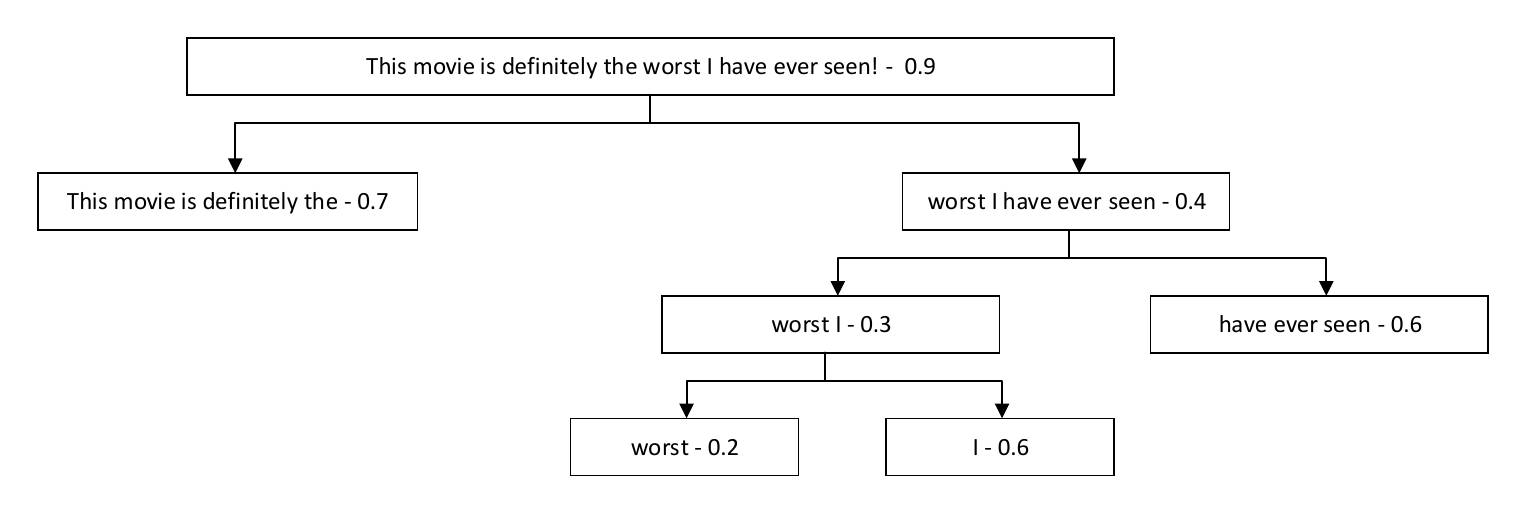}
    \caption{Visualization of Binary Tree leveraged to store the probabilities returned from \textit{BinarySelect}. Note that the values indicate probability of target class. After the root node, the probabilities are calculated by removing the text at that node from the original text.}
    \label{fig:bsstruct}
\end{figure*}

\subsection{Greedy Select}
%The greedy search algorithm systematically explores the impact of individual word removals from a given sentence, seeking to identify the word whose absence induces the most significant shift in sentiment prediction. This iterative process allows for the identification of locally optimal word candidates that wield substantial influence over the sentiment classification. 
Greedy select and variations have been strongly leveraged by previous black-box attacks \cite{ren-etal-2019-generating, li-etal-2020-bert-attack, hsieh-etal-2019-robustness, formento-etal-2023-using, garg2020bae,gao2018blackbox, jin2020bert, Li2019Textbugger, rusert2022robustness}.
This method deletes one word at a time (with replacement) and checks the change in classifier probability. More formally, let $f$ represent the classifier, and $f(X)$ represent the classification score (i.e. probability). For each word $w_i$ in $X$, greedy select removes the word and finds $\Delta_i$:
\begin{equation}
    \Delta_i = f(X) - f(X//w_i)
\end{equation}
which is the change in probability of a target class when $w_i$ is removed. The word with the highest drop in target class probability is selected as the word to replace. Note here we define it with deletion, however, some of the attacks replace the word with masks instead \cite{li-etal-2020-bert-attack}. Furthermore, other variations include scores of different portions of the text \cite{gao2018blackbox}. Nevertheless, all the related methods remove one word at a time to find the greatest drop. This results in at least $n$ queries.

\subsection{\textit{BinarySelect}}\label{sect:bs}
\textit{BinarySelect} (Algorithm in Appendix \ref{sect:bsalg}) adopts a more systematic and targeted approach. By dividing the sentence and evaluating classification score changes, it greatly narrows down the search space. This technique builds off of the binary search algorithm by viewing the texts as larger segments to search. %This greatly reduces the search complexity while maintaining a high level of precision. 

In \textit{BinarySelect}, the text is continuously partitioned into two segments until a segment containing a single word is reached. At each step, the method evaluates the impact of excluding each segment on the classifier's output probability and selects the segment that results in the greatest drop in probability. More formally, let $f$ represent the classifier, and $f(X)$ represent the classification score (i.e., probability). Given an input text $X$, \textit{BinarySelect} partitions $X$ into two segments, $X_1$ and $X_2$, such that $X_1 \cup X_2 = X$ and $X_1 \cap X_2 = \emptyset$. The method calculates the probabilities $f(X_1)$ and $f(X_2)$ by excluding $X_2$ and $X_1$, respectively, from the original text $X$. The difference in probability for each segment is computed as:
\begin{equation}
\Delta_i = f(X) - f(X_i)
\end{equation}
where $\Delta_i$ represents the change in the probability of a target class when segment $X_i$ is excluded from the input text $X$. The segment with the highest probability drop of the target class, is processed further. If the segment is a single word, then this word is chosen as the most influential word. If it is not, then the process repeats with the segment becoming the next text to be divided in two. 

\subsection{\textit{BinarySelect} - Retaining Memory}\label{sect:bsstruct}

For GreedySelect, repeating the word selection stage is simple since all probability drops are calculated in the first pass through. However, \textit{BinarySelect} only has a score for a single word in the text. To avoid additional queries, we leverage a binary tree structure to keep track of which segments the algorithm has generated scores for. 

A visualization of this can be seen in Figure \ref{fig:bsstruct}. This structure is continually updated as new segments are queried against the classifier (described in Section \ref{sect:bs}). During the selection step, \textit{BinarySelect} explores the tree path with the greatest drop in probability which hasn't been fully explored.

\section{Theoretical Performance}
We examine the theoretical performance of \textit{BinarySelect} by examining 3 cases:

\noindent \textbf{Best Case:} In the best case, only one word is needed to be found. In this case, \textit{BinarySelect} takes at most $\text{log}_2(n) * 2$ queries. This is because it takes $\text{log}_2(n)$ splits to reach a single word and each split requires 2 queries to guide the method. This value is less than GreedySelect which takes $n$ queries for even 1 word, since it needs to remove every word and test the probability changed. 

\noindent \textbf{Average Case:} Since each dataset and classifier can rely on a different number of words for classification, it can be difficult to know how many words are relied upon for classification on average. We estimate this value based on previous research. BERT-Attack \cite{li-etal-2020-bert-attack}, reports the percentage of a text that is perturbed during the attack for IMDB and AG News (Section \ref{sect:experimentalsetup}). This percentage is 4.4 for IMDB and 15.4 for AG News. We use this to estimate the number of words changed on average (percentage X average text length). For IMDB this results in an average of 9.5 words being changed (average text length of 215) and 6.6 words for AG News (average text length of 43). This means to find the words, GreedySelect would need the average number of words as queries (215 and 43).

For \textit{BinarySelect} it is non-deterministic, since any query after the first, will leverage the binary structure (Appendix \ref{sect:bsstruct}). The first query is again $\text{log}_2(n) * 2$ queries. For the second query the first split and query is not needed as it was estimated previously. In the worst case scenario, the half of the tree that wasn't explored contains the next largest drop in probability and thus it is expanded. This means, in the worst case, the second query requires $\text{log}_2(n/2) * 2$ queries. We can follow this worst case scenario as a basis to estimate the number of queries needed for \textit{BinarySelect}. This results in a value of $\text{log}_2(n) * 2 + \text{log}_2(n/2) * 2 + \text{log}_2(n/4) * 2 + ... + \text{log}_2(n/(2^k-1)) * 2$. Note that when $2^k-1$ is larger than $n$, then we cap the value at 1, since at the lowest level, we need 2 queries for the two words being split. 

For IMDB, with an average of 10 (9.6) word changes this results in 72 queries. For AG News, with an average of 7 (6.6) word changes, this value is 37. Note that this value would only be lower if the next most probable words are in the already explored structure. We can then see that even in the average (worst-case) \textit{BinarySelect} requires less queries to find the same number of words as \textit{GreedySelect}. We can also note a stronger performance increase in longer texts.

\noindent \textbf{Worst Case:} In the extreme worst case, we need to find the probabilities of every word in the input text. For GreedySelect, this is again equal to $n$. However, for \textit{BinarySelect}, this is much greater as it will make $n$ queries, as well as each split level queries. This results in $n + \sum_{i=1}^{\text{log}_2(n)} n/(2^i)$. This scenario shows a disadvantage of \textit{BinarySelect} and shows that it is not a permanent fix for \textit{GreedySelect}, especially in scenarios where replacements are needed for a large percentage of the text.

\section{Validation of BinarySelect}

\begin{table}[]
    \centering
    \footnotesize
    \begin{tabular}{c|c|c}
         Token \# & AG News & IMDB\\\hline
         1 & 12.5 & 17.2\\
         2 & 17.9 & 25.0\\
         3 & 21.6 & 30.9\\
         4 & 24.4 & 35.8\\
         5 & 26.7 & 40.0\\
         6 & 29.0 & 44.0\\
         7 & 30.9 & 47.6\\
         8 & 32.7 & 51.0\\
         9 & 34.5 & 54.2 \\
         10 & 35.9 & 57.0 \\\hline
         GS & 39.5 & 230.6 \\\hline
    \end{tabular}
    \caption{Average Queries to find the $k$ top words for BS. Since GS requires all words to be checked, the number of queries is the same for all 10.}
    \label{tab:avg_queries}
\end{table}
\begin{figure}
    \centering
    
    \includegraphics[width=0.5\textwidth]{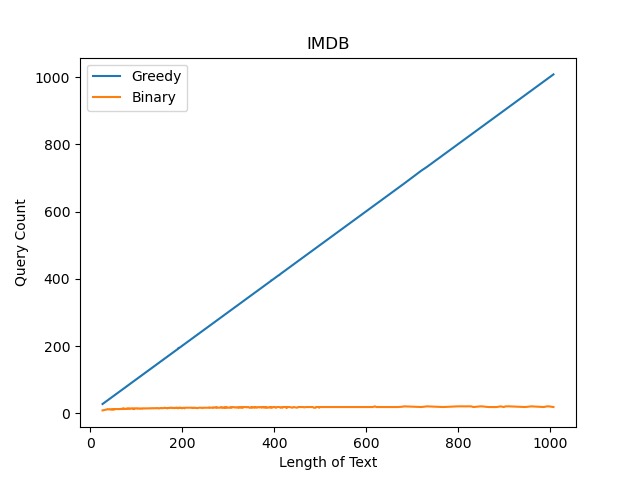}
    \caption{Number of queries required to find a word leveraged by the classifier. GreedySelect's queries increase linearly, while \textit{BinarySelect} shows a log trend.}
    \label{fig:validation}
\end{figure}
\iffalse
\begin{table*}[]
    \centering
    \footnotesize
    \begin{tabular}{c|c|c|c|c|c|c|c|c|c|c||c}
          & 1 & 2 & 3 & 4 & 5 & 6 & 7 & 8 & 9 & 10 & GreedySelect  \\\hline
         AG News & 12.5 &  17.9 & 21.5 & 24.4 & 26.7 & 29.0 & 30.9 & 32.7 & 34.5 & 35.9 & 39.5\\
         IMDB & 17.2 & 25.0 & 30.9 & 35.8 & 40.0 & 44.0 & 47.6 & 51.0  & 54.2 & 57.0 & 230.6 \\\hline
    \end{tabular}
    \caption{Avg. Queries to find $k$ top words for BS. Since GS requires all words to be checked, the number of queries is the same for all 10.}
    \label{tab:avg_queries}
\end{table*}
\fi
\subsection{Verification of Theoretical Performance}
We validate the theoretical performance by running the binary select algorithm on 1000 AG News and IMDB (Section \ref{sect:experimentalsetup}) examples, using a fine-tuned ALBERT \cite{devlin-etal-2019-bert} classifier for feedback. We retrieve the top 10 words and note how many queries were required for each word. The results can be seen in Table \ref{tab:avg_queries}. 
We find that on average the number of queries is less than the estimated amount in the theoretical case. For AG News, the estimated average value was 37, while the found average was 30.9 (7 words). Similarily, for IMDB, the estimated value was 72 and the found value was 57. This is because in the estimated value, we looked at the worst case of the average case where an unexplored branch of the binary structure was explored every query. However, this was not the case in the experiments. In both cases, the needed queries is lower than what is needed for the GreedySelect for all 10 queries.  

%\subsection{Pilot Study: Validation of \textit{BinarySelect}}
We further validate \textit{BinarySelect}'s advantage by examining how many queries the method needs to find the most influential word. We do this for the 1000 IMDB examples and note the length of the text and number of queries required to find the most influential word. These results can be found in Figure \ref{fig:validation}. We can observe that the GreedySelect has a linear trend, while the \textit{BinarySelect} follows a log trend, this indicates a significant difference in computational complexity between the two methods. Specifically, it suggests that as the length of the input sentence increases, the query count for the GreedySelect tends to increase linearly. In contrast, the \textit{BinarySelect} demonstrates a more consistent log query count regardless of sentence length. This observation emphasizes the efficiency advantage of \textit{BinarySelect} over GreedySelect, particularly when dealing with longer input texts. The log trend  for \textit{BinarySelect} suggests that its computational requirements remain relatively stable and independent of input size, which can be a highly advantageous characteristic in practical applications.

\begin{table}[]
    \centering
    \footnotesize
    \begin{tabular}{c|c|c|c}
          & & BinarySelect & Random \\\hline
          & \multicolumn{3}{c}{Position of 1st GS Token}  \\\hline
          \parbox[t]{2mm}{\multirow{3}{*}{\rotatebox[origin=c]{90}{AG}}} & Average & 2.3 & 5.0 \\
          & Median & 1 & 5 \\
          & Not Found & 255 & 647 \\\hline
           \parbox[t]{2mm}{\multirow{3}{*}{\rotatebox[origin=c]{90}{IMDB}}} & Average & 2.9 & 5.1 \\
          & Median & 2 & 4 \\
          & Not Found & 583 & 921 \\\hline\hline
          & \multicolumn{3}{c}{Num. Overlaps with GS top 10}\\\hline
          \parbox[t]{2mm}{\multirow{3}{*}{\rotatebox[origin=c]{90}{AG}}} & Average & 5.7 & 3.4 \\
          & Median & 6 & 3 \\
          & None & 4 & 15 \\\hline
          \parbox[t]{2mm}{\multirow{3}{*}{\rotatebox[origin=c]{90}{IMDB}}} & Average & 3.7 & 1.7 \\
          & Median & 4 & 1 \\
          & None & 110 & 345 \\\hline
    \end{tabular}
    \caption{Comparison between the top 10 words found by GreedySelect (GS). Position of 1st refers to which position the top GS token is found by the respective method (low values desired). Num. Overlaps refers to the number the top 10 GS tokens appear in the top 10 found by the other method (higher values desired).}
    \label{tab:agreement}
\end{table}

\subsection{Agreement between BinarySelect and GreedySelect}
It is not sufficient for BinarySelect to find influential words more efficiently than GreedySelect, we also need BinarySelect to find words that are truly relevant. To verify this, we run 2 more experiments which measure on if GreedySelect and BinarySelect agree on the most influential words. 

First, we take the top word given by GS and note which influential position it was given by BS. If GS and BS always agree on the most influential word, then that position will be 1. We examine the top 10 words found by BS for both the AG News and IMDB examples. We also make a random baseline which randomly choose 10 words in the input texts. The results can be found in Table \ref{tab:agreement}. We find that for AG News, the most influential word found by GS appears in the 2.3 position on average and the 1 position for a median value. These are much lower than the random baseline which the position is 5 on average and median. In 255 of the texts, the top word of GS is not found in the top 10 BS list. This is also much lower than the random baseline which 647 texts do not include the word. For IMDB, these values are slightly larger (since the lengths of IMDB texts are much longer), with averages of 2.9 and 5.1 for BS and random respectively and median values of 2 and 4 respectively. 

Second, we look at how many words in the top 10, BS and GS agree on. For the same examples, we note how many of the BS words occur in the GS list. For AG News, we find that the BS list has 5.7 of the same words on average (median of 6), which is more than the random at an average of 3.4 (median of 3). For IMDB these values are slightly lower (again due to IMDB text's lengths), on average the BS list contains 3.7 words (median of 4) which is still higher than the random baseline average of 1.7 (median of 1). 

We see that though BS and GS do not completely agree on the most influential words, there is ample overlap between the two methods. Note that these experiments can only measure agreement and not which is the most ``effective'' in downstream tasks. To verify this, we leverage both BS and GS in a common setting, Adversarial Attacks (Section \ref{sect:adversarialattack}).

% V This doesn't have anything to do with our validation experiment, also sentiment analysis is not all three datasets -Jon

%In each iteration, the input text is divided into two segments, and sentiment analysis is performed on each segment to calculate the drop in sentiment probability. By comparing the drop in probability for both the segments and updating the search range accordingly, the algorithm efficiently narrows down the search to identify the position of the most influential word. This approach optimizes query count, facilitating the generation of adversarial examples in text-based applications. The algorithm's performance is evaluated on the IMDB dataset, which comprises movie reviews labeled with sentiment polarity (0 or 1). For each example in the dataset, the algorithm identifies the most influential word and calculates the change in sentiment score, providing insights into its effectiveness in generating adversarial examples.

%\textbf{WE WILL ALSO NEED TO ADD THE HUMAN VERIFICATION IF APPLICABLE}

%\subsection{Observations}

\section{Testing \textit{BinarySelect} in Adversarial Attacks}\label{sect:adversarialattack}

As noted, GreedySelect is widely leveraged by many black-box attacks \cite{ren-etal-2019-generating, li-etal-2020-bert-attack, hsieh-etal-2019-robustness, formento-etal-2023-using, garg2020bae,gao2018blackbox, jin2020bert, Li2019Textbugger, rusert2022robustness}. However, its need to examine every token in a text can slow down the attack algorithm greatly, which causes barriers for researchers with low resources. Thus, \textit{BinarySelect} may be a strong replacement to decrease the overall number of queries required per attack. 

As a reminder, in adversarial attacks, the goal is to create input examples that are understood (by humans) similarly to the original ones but lead to incorrect classifier predictions. This is often accomplished by modifying one word of a text at a time and checking against the classifier noting changes in probabilities. Once the modified text causes the classifier to fail, the attack ends. Note that the attacker must also aim to maintain the semantic integrity of the text to keep meaning.

\subsection{Attack Description}
To test the feasibility of \textit{BinarySelect} in attacks, we create a similar attack framework to previous research. Our attack consists of two steps, word selection and word replacement:

1. Word Selection - the position of the word which the classifier relies on the most for classification is chosen to be replaced. Either \textit{BinarySelect} or GreedySelect is used to find this position.

2. Word Replacement - the word at the selected position is replaced. We query WordNet for the selected word's synonyms. Each synonym is tested and the synonym which causes the classifier to fail or causes the greatest drop in target class probability is chosen. If the classifier does not fail with this replacement, the process repeats with Word Selection. However, this time the previous modified position is excluded as a candidate. 

Note that this word replacement step is similar to PWWS \cite{ren-etal-2019-generating}. A more advanced replacement step would generate a stronger attack, however, we choose this simple replacement step to place the focus  on the selection algorithms.

\subsection{Restriction to $k$ Words}
To further improve the efficiency of the attack, we add the option of restricting the attack to modify at most $k$ words. As $k$ increases the attack effectiveness will naturally increase but the number of queries required will increase as well. Additionally, as more words change, the semantic integrity starts to weaken. This $k$ is another useful tool for allowing researchers with lower resources to control effectiveness versus efficiency. We explore the effect of $k$ in Section \ref{sect:choosingk}.

\begin{table*}[]
    \centering
    \footnotesize
    \begin{tabular}{c|c||cc||cc||cc||cc||cc||}
          &  & \multicolumn{2}{|c||}{Albert} & \multicolumn{2}{|c||}{Distilbert} & \multicolumn{2}{|c||}{BERT} &  \multicolumn{2}{|c||}{Roberta} &  \multicolumn{2}{|c||}{LSTM}\\\hline
          &  & GS & BS & GS & BS & GS & BS & GS & BS & GS & BS
         \\\hline

         % yelp 
          \parbox[t]{2mm}{\multirow{4}{*}{\rotatebox[origin=c]{90}{Yelp }}}& Original Acc. & \multicolumn{2}{|c||}{99.8} & \multicolumn{2}{|c||}{95.2} & \multicolumn{2}{|c||}{99.5} &  \multicolumn{2}{|c||}{98.3} & \multicolumn{2}{|c||}{94.7}\\ 
         & Attack Acc. & 43.5 & 51.7 & 31.1 & 46.6  & 47.2 & 52.6 & 54.5 & 65.3 & 10.9 & 32.2\\\cline{2-12}
         & Avg. Queries & 217 & 150 & 208 & 141 & 222 & 150  & 239 & 172 & 181 & 119\\
         & Avg. Q's (Success) & 156 & 93 &  162 & 93 & 150 & 100 & 160 & 107 & 173 & 91\\\hline\hline

        % imdb 
          \parbox[t]{2mm}{\multirow{4}{*}{\rotatebox[origin=c]{90}{IMDB}}}& Original Acc. & \multicolumn{2}{|c||}{97.7} & \multicolumn{2}{|c||}{96.8} & \multicolumn{2}{|c||}{97.9} &  \multicolumn{2}{|c||}{97.6} & \multicolumn{2}{|c||}{84.8}\\ 
         & Attack Acc. & 51.8 & 66.9 & 37.2 & 58.2  & 54.4 & 70.0 & 55.0 & 72.5 & 25.4 & 52.9\\\cline{2-12}
         & Avg. Queries & 318 & 172 & 305 & 156 & 317 & 173 & 332 & 182 & 274 & 136 \\
         & Avg. Q's (Success) & 273 & 106 & 265 & 99 & 269 & 110 & 275 & 113 & 262 & 96\\\hline\hline

        % ag news 
          \parbox[t]{2mm}{\multirow{4}{*}{\rotatebox[origin=c]{90}{AG News}}}& Original Acc. & \multicolumn{2}{|c||}{98.8} & \multicolumn{2}{|c||}{97.4} & \multicolumn{2}{|c||}{99.6} &  \multicolumn{2}{|c||}{99.2} & \multicolumn{2}{|c||}{93.1}\\ 
         & Attack Acc. & 46.2 & 48.2 & 60.7 & 62.8 & 62.6 & 64.4 &  55.9 & 58.3 & 43.5 & 47.7 \\\cline{2-12}
         & Avg. Queries & 111 & 111 & 121 & 124 & 125 & 127 &  119 & 121 & 104 & 112 \\
         & Avg. Q's (Success) & 84 & 76 & 92 & 86 & 89 & 84 & 86 & 82 & 84 & 85\\\hline\hline
         
    \end{tabular}
    \caption{Adversarial Attack Results when $k=15$. ``Original Acc.'' is the original accuracy of the model, ``Attack Acc.'' is the model accuracy on the text modified by the attack. ``Avg. Queries'' is the average number of queries used, ``Avg. Q's (Success)'' are the number of queries used for successful attacks. GS - GreedySelect, BS - \textit{BinarySelect}.}
    \label{tab:attackResults}
\end{table*}

\section{Experimental Setup}\label{sect:experimentalsetup}
To evaluate \textit{BinarySelect} in the adversarial attack setting, we run the attack (Section \ref{sect:adversarialattack}) on 5 classifiers across 3 datasets\footnote{The majority of attacks are run on Google Colab and Kaggle which use NVidia K80 GPUs. Each attack combination took roughly 40 minutes.}. For space, the datasets and classifiers are described in detail in Appendix \ref{sect:experimentaldetails}.

\subsection{Metrics}\label{sect:metrics}
We use the following metrics to evaluate \textit{BinarySelect} in the attack:

1. Accuracy - We measure the accuracy of each model before and after the attack for both GreedySelect (GS) and \textit{BinarySelect} (BS). This helps measure the strength of the attack for each. 

%2. Attack Success Rate (ASR) (Equation \ref{eq:asr} - Metric used in previous research \cite{}. Helps contrast the drop in accuracy with the original accuracy. 
%\begin{equation}\label{eq:asr}
%    \text{ASR} = \frac{\text{Original}_\text{Acc.}- \text{Attacked}_\text{Acc.}}{ \text{Original}_\text{Acc.}}
%\end{equation}

2. Average Queries - To measure the queries saved by using \textit{BinarySelect}, we measure the number of queries needed for an attack on average. These queries indicate how many calls to the classifier are needed.

3. Average Queries when Successful -  Similar to average queries, but in the cases when the attack is successful. \textit{BinarySelect} will naturally suffer when more of the text is explored, which is what happens in failed attacks. This measurement shows an ideal case for the attack. 

4. Effectiveness Differential Ratio (EDR) - To measure the tradeoff between Attack Success Rate (Equation \ref{eq:asr})(ASR) and Average Queries, we propose EDR (Equation \ref{eq:edr}), which contrasts the percentage change in ASR (Equation \ref{eq:asrchange}) with the percentage change in Average Queries (Equation \ref{eq:querychange}). We use this measure to help explore how $k$ affects BS versus GS (Section \ref{sect:choosingk}). 

\begin{equation}\label{eq:asr}
    \text{ASR} = \frac{Original_{Acc.}- Attack_{Acc.}}{ Original_{Acc.}}
\end{equation}

\begin{equation}\label{eq:querychange}
    \text{Query}_\text{Diff} = \frac{\text{Queries}_\text{Greedy} -\text{Queries}_\text{Binary}}{\text{Queries}_\text{Greedy}}
\end{equation}

\begin{equation}\label{eq:asrchange}
    \text{ASR}_\text{Diff} = \frac{\text{ASR}_\text{Binary} - \text{ASR}_\text{Greedy}}{\text{ASR}_\text{Greedy}}
\end{equation}

\begin{equation}\label{eq:edr}
    \text{EDR} = \text{ASR}_\text{Diff} + \text{Query}_\text{Diff}
\end{equation}

%\subsection{Dataset}
%Our study begins with the careful curation of datasets for evaluation. The cornerstone of our investigation is the IMDB dataset, a widely acknowledged benchmark in sentiment analysis. This corpus comprises a diverse array of movie reviews, each annotated with sentiment labels. This diverse selection aims to capture the nuances and challenges associated with sentiment analysis in real-world scenarios.
\section{Results}\label{sect:results}
\begin{figure*}
    \centering
    \footnotesize
    \begin{subfigure}[b]{0.67\columnwidth}
        \fbox{\includegraphics[width=0.95\textwidth]{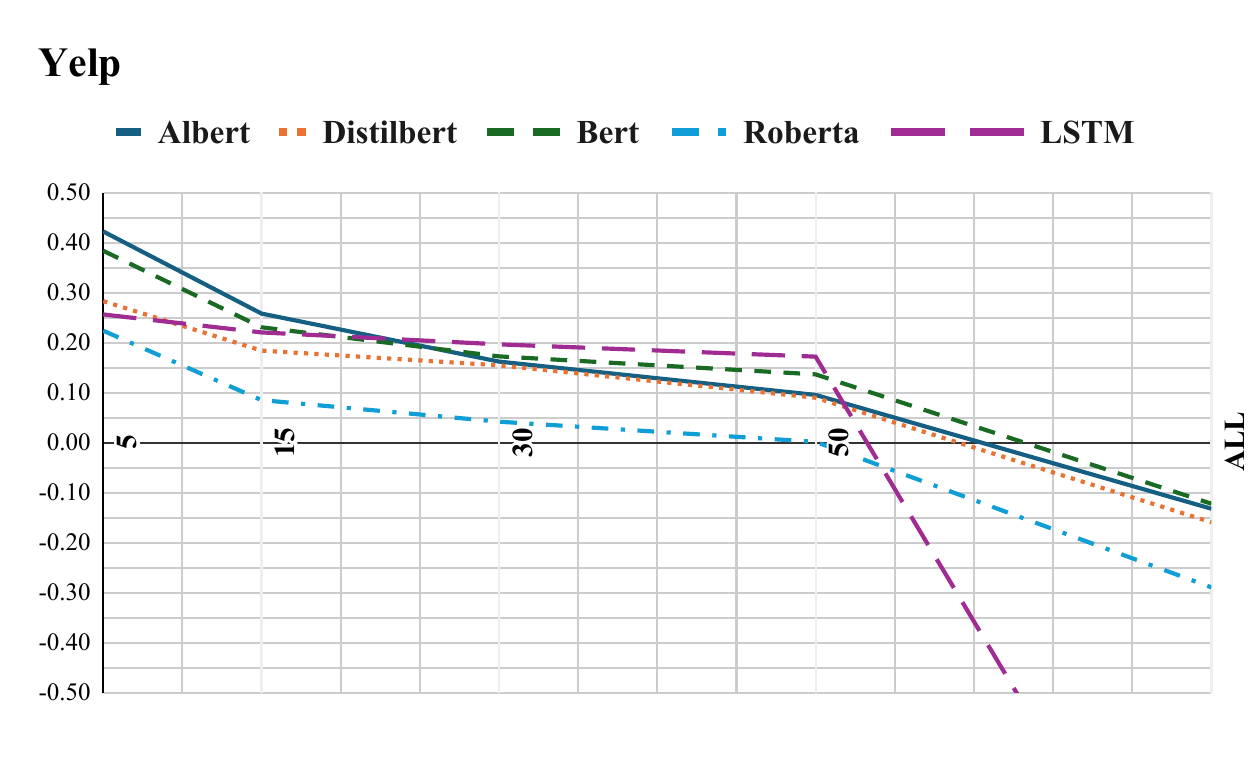}}
    \end{subfigure}
    \begin{subfigure}[b]{0.67\columnwidth}
        \fbox{\includegraphics[width=0.95\textwidth]{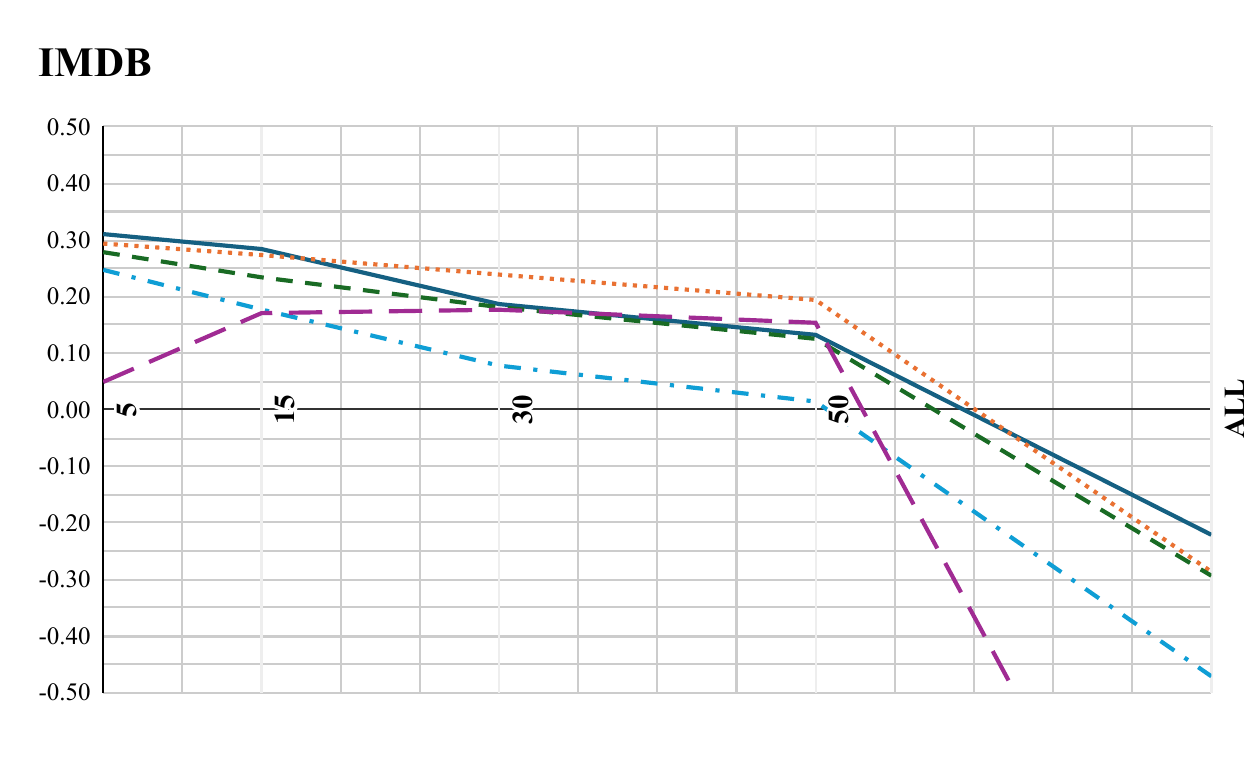}}
    \end{subfigure}
    \begin{subfigure}[b]{0.67\columnwidth}
        \fbox{\includegraphics[width=0.95\textwidth]{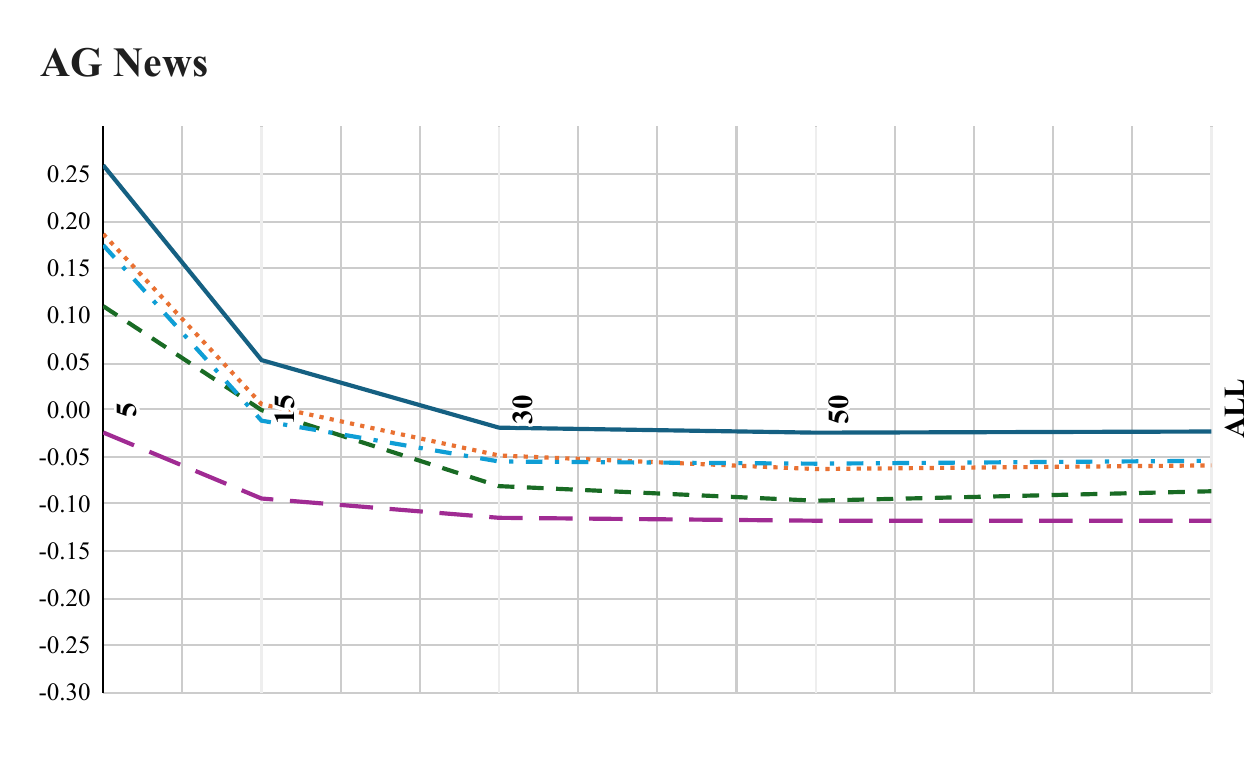}}
    \end{subfigure}
    \caption{ Effect of $k$ Values on EDR (Equation \ref{eq:edr}) for the successful attacks. Positive values indicate a better trade-off between reduction in queries versus loss of accuracy drop for BS.}
    \label{fig:kcharts}
\end{figure*}
\begin{table*}[]
    \centering
    \footnotesize
    \begin{tabular}{c||cc||cc||cc||cc||cc}
         & \multicolumn{2}{c|}{$k=5$} & \multicolumn{2}{c|}{$k=15$} & \multicolumn{2}{c|}{$k=30$} & \multicolumn{2}{c|}{$k=50$} & \multicolumn{2}{c}{$k=ALL$}  \\\hline
         & GS & BS & GS & BS & GS & BS & GS & BS & GS & BS\\\hline
         Orig Acc.& \multicolumn{10}{c}{85.8} \\
         Attack Acc. & 47.8 & 56.0 & 29.2 & 38.5 & 24.1 & 32.0 & 22.8 & 28.9 &  22.9 & 27.9 \\\hline
         Avg. Queries & 108 & 31 & 112 & 50 & 117 & 68 & 122 & 85 & 135 & 133 \\
         Avg. Q's (Success) & 101 & 23 & 108 & 34 & 111 & 42 & 113  & 47 & 113 & 49 \\\hline
    \end{tabular}
    \caption{Character Level attack using BS and GS against canine-s finetuned on SST2 data. }
    \label{tab:sst2}
\end{table*}

The main results for our attack experiments can be found in Table \ref{tab:attackResults}. For each classifier, we compare the GreedySelect (GS) and \textit{BinarySelect} (BS). We use a $k$ value\footnote{Tables for other $k$ can be found in Appendix \ref{app:kresults}} of 15, which means the attack was limited to replacing 15 words in a text at most (a further exploration of $k$ values can be found in Section \ref{sect:choosingk}). The first three metrics described in \ref{sect:metrics} are shown. The following observations are made:

\noindent \textbf{\textit{BinarySelect} reduces the number of queries greatly, with some drop in attack effectiveness.} When examining Table \ref{tab:attackResults}, we see drops in query amounts  for both IMDB and Yelp datasets. Focusing on the Albert classifier, we see a 31\% difference in queries between GreedySelect (217) and \textit{BinarySelect} (150). This drop in queries causes a slight drop in attack effectiveness of 16\%. Similarly for IMDB, we see a larger difference in queries. Specifically, \textit{BinarySelect} causes a 46\% reduction in queries compared to GreedySelect, with a 23\% drop in attack effectiveness.  If we focus on the number of queries for successful attacks, then this increases to a 54\% reduction in queries for the same 23\% effectiveness tradeoff. Hence, we see a stronger positive effect in query reduction compared to attack effectiveness.

\noindent \textbf{\textit{BinarySelect} is less effective on shorter texts.} For AG New, we see similar results between GreedySelect and \textit{BinarySelect}. Both the query numbers and accuracy are within a few points of each other. A main reason is that AG News contains shorter texts on average (43 words) compared to Yelp (157) and IMDB (215). This means \textit{BinarySelect} will save less queries with each search for AG News compared to Yelp and IMDB. Nonetheless, we see that in the extended case, \textit{BinarySelect} still achieves comparable performance to GreedySelect.

\noindent \textbf{\textit{BinarySelect} strongest effect is demonstrated on Yelp Dataset.}
The results on the Yelp dataset show a clear instance in the strength of \textit{BinarySelect}. We see a 32\% reduction in queries for BERT, a 31\% reduction for Albert, and a 32\% reduction for Distilbert. The effectiveness of the attack is at a lower rate as well, for example an 10\% drop for BERT and 16\% for albert. Distilbert seems to be an exception with a larger drop in effectiveness. These results demonstrate the true potential for \textit{BinarySelect}.

\noindent \textbf{Successful attacks cause a greater reduction in attack queries.} 
When examining the average queries for the succesful attacks, we see a greater reduction in query amount on average. For IMDB, the reductions increase from 46\% (Albert), 45\% (XLNet), and 45\% (RoBERTa) to 61\% (Albert), 60\% (XLNet), and 59\% (RoBERTa). This indicates that if a more successful replacement step is chosen, then the algorithm will increase in effectiveness. 

In all observations we see a clear reduction in query amounts with a lesser reduction in attack effectiveness. This helps highlight the tradeoffs of \textit{BinarySelect}. Future work would implement the algorithm with more effective replacement/modification steps to extend \textit{BinarySelect} to its full potential.

\section{Choosing an Effective $k$} \label{sect:choosingk}
The main attack results use $k=15$, however, the chosen $k$ will impact both attack effectiveness and queries needed. Different datasets will benefit from different $k$. To investigate this, for each dataset and classifier, we test $k = \{5, 15, 30, 50, \text{ALL}\}$ where ``ALL'' imposes no restrictions on the number of words to replace. Figure \ref{fig:kcharts} shows the effect of different $k$ for successful attacks, measured with EDR (Equation \ref{eq:edr}). As $k$ increases the query amount increases and the accuracy of the targeted classifier decreases. This causes a better trade off of EDR at lower $k$. The optimal $k$ will minimize the accuracy and minimize the amount of queries We see that $k={15,30}$ offers a balance for Yelp and IMDB, but $k=5$ would be better for AG News. These results demonstrate the need for testing different $k$ for different tradeoff goals.

\iffalse
\begin{figure}
    \centering
    \begin{subfigure}[b]{0.95\columnwidth}
        \centering
        \fbox{\includegraphics[width=\textwidth]{images/accuracyk.pdf}}
    \end{subfigure}
    \begin{subfigure}[b]{0.95\columnwidth}
        \centering
        \fbox{\includegraphics[width=\textwidth]{images/avgqs.pdf}}
    \end{subfigure}
    \begin{subfigure}[b]{0.95\columnwidth}
        \centering
        \fbox{\includegraphics[width=\textwidth]{images/avgqssuccess.pdf}}
    \end{subfigure}
    \caption{Effect of $k$ Values on accuracy and average queries for Albert on the Yelp dataset.}
    \label{fig:kcharts}
\end{figure}
\fi

\section{Verification: Character-level Attack}
We verify the main results of \textit{BinarySelect} by extending the experiments to character-level attacks. Specifically, we target a character-level model \cite{canine-google}, fine-tuned on the SST2 (Stanford Sentiment Treebank, contains movie reviews labeled for sentiments) dataset. We leverage BS and GS to choose which character to modify, and use the ECES unicode replacement from VIPER \cite{eger-etal-2019-text} to replace each chosen character. We run the attack for $k=5$ and $k=ALL$ on the validation set (872 instances). The results can be found in Table \ref{tab:sst2}.

We observe results consistent (or better) with the word level attacks (Section \ref{sect:results}). For the lower $k$, we see \textit{BinarySelect} outperform GreedySelect, while for large $k$, we see similar attack effectiveness and similar query counts. However, again for the successful cases, BS strongly outperforms GS, which further points to the potential strengths of BS.
\section{Further Analysis: Combining \textit{BinarySelect} and GreedySelect}
We see that in our results (Section \ref{sect:results}), there exists a tradeoff between using \textit{BinarySelect} and GreedySelect. Furthermore, the chosen $k$ greatly affects this tradeoff. Specfically, we see a better EDR with lower $k$. This means, that \textit{BinarySelect} is a better choice for texts which require less word changes. To further examine this, we imagine an \textbf{oracle} model which knows how many words need to be changed for an attack to succeed (We compare the modified texts by \textit{BinarySelect} to determine the number). The oracle can leverage the strengths of both \textit{BinarySelect} and GreedySelect. If the number of words to be changed is less than $j$, then \textit{BinarySelect} is used, otherwise GreedySelect is used. We compare this \textbf{oracle} model with the previous results ($k$ = ALL) for DistilBert on IMDB with various $j$ in Table \ref{tab:oracle}.

\begin{table}[]
    \centering
    \footnotesize
    \begin{tabular}{c|c|c}
         Model & Attack Acc. & Avg. Q's\\\hline
         GS & 3.4 & 407 \\
         BS & 3.8 & 526 \\\hline\hline
         \multicolumn{3}{c}{Oracle}\\\hline
         $j$ <= 5 & 3.4 & 369 \\
         $j$ <= 15 & 3.4 & 346 \\
         $j$ <= 30 & 3.4 & 341 \\
         $j$ <= 50 & 3.4 & 358 \\
         $j$ > 50 & 3.8 & 575 \\ \hline\hline
    \end{tabular}
    \caption{Combination results for GS and BS on IMDB data for DistilBert. Oracle knows how many words need to be perturbed for an attack and uses BS for texts less than $j$ and GS for those more than $j$.}
    \label{tab:oracle}
\end{table}

As can be observed, the oracle is able to achieve GS's lower accuracy, with much lower queries overall by leveraging both BS and GS effectively. This \textbf{oracle} model is the ideal to strive for, however, we do not automatically know how many words will need to be changed for a target model to fail. We run a preliminary experiment to determine if confidence score can be utilized as an oracle (Appendix \ref{sect:confidencemodel}), but find it does not perform as well as the oracle. Future work will further investigate discovering this automatic oracle to most effectively utilize BS and GS.
\subsection{Confidence Model to Emulate \textbf{Oracle}}\label{sect:confidencemodel}
\begin{table}[]
    \footnotesize
    \centering
    \begin{tabular}{c|c|c}
         Model & Attack Acc. & Avg. Q's\\\hline
         GS & 3.4 & 407 \\
         BS & 3.8 & 526 \\\hline\hline
         \multicolumn{3}{c}{Oracle}\\\hline
         $j$ <= 5 & 3.4 & 369 \\
         $j$ <= 15 & 3.4 & 346 \\
         $j$ <= 30 & 3.4 & 341 \\
         $j$ <= 50 & 3.4 & 358 \\
         $j$ > 50 & 3.8 & 575 \\ \hline\hline
         \multicolumn{3}{c}{Confidence Model}\\\hline
         Score <= $\text{Avg}_5$ & 3.4 & 422 \\
         Score <= $\text{Avg}_15$ & 3.4 & 419 \\
         Score <= $\text{Avg}_30$ & 3.4 & 422 \\
         Score <= $\text{Avg}_50$ & 3.4 & 422 \\
         Score > $\text{Avg}_50$ & 3.4 & 511 \\
    \end{tabular}
    \caption{Combination results for GS and BS on IMDB data for DistilBert. Oracle knows how many words need to be perturbed for an attack and uses BS for texts less than $j$ and GS for those more than $j$. }
    \label{tab:confidencemodel}
\end{table}

As a preliminary, we look at the confidence (probabilities) of the classifier on the original text as an indicator. When binning the average confidence scores against the number of word changes, we find a slight pattern of increase: [ 5 - 93.77, 15 - 96.27, 30 - 97.12, 50 - 97.46, ALL - 97.67]. However, in the larger changes, there exists slight differences in the average confidence scores. Nonetheless, we try a secondary model which uses the noted confidence scores to determine if BS or GS is used. Similarly to \textbf{Oracle}, if the original confidence score is less than the average confidence for a bin, then BS is used, otherwise GS. Note that this is still part oracle, as the average confidence scores would not be known. These results are found in Table \ref{tab:confidencemodel}. As can be observed, the confidence model performs better than BS but similarly to GS, and not as well as the \textbf{Oracle} model. This means the confidence score alone is not adequate to determine when to use BS versus GS. 
\section{Conclusion}
\textit{BinarySelect} shows a strong promise to increase efficiency of attack research and other related domains. Specifically, we found that \textit{BinarySelect} is able to find a word relevant to a classifier in $\text{log}_2(n) * 2$ steps. This is much more efficient that GreedySelect and its variants which take $n$ (or more) queries to produce a word. 

We further tested \textit{BinarySelect} in the downstream task of adversarial attacks. To keep focus on the selection method, we combined it with a WordNet replacement method. We found a viable tradeoff between query reduction and drop in attack effectiveness. For BERT on the Yelp dataset, \textit{BinarySelect} takes 32\% (72) less queries than GreedySelect with only a 10\% (5 point) drop in attack effectiveness. Furthermore by including the choice for a $k$, we introduced more control to the researcher. We further verified this on a character-level attack. Finally, we showed the potential for ideal method that combines \textit{BinarySelect} and GreedySelect, however, it is left to future research to fully solve this problem.

GreedySelect's frequent usage in multiple emi-nent attacks is resource draining.  \textit{BinarySelect} is effective in giving low-resource researchers the ability to be apart of this domain, allowing the best ideas a chance to be realized.
\section{Limitations}
Here we note limitations of our study for future researchers and users to consider:

1. \textbf{Stronger Replacement Steps Exist for Attacks} - Our algorithm was limited in measurement due to leveraging WordNet alone as replacement. Other attack research has leveraged transformer models such as BERT to give more relevant suggestions for replacement. This could have resulted in earlier stopping in the attack due to better replacement choices. However, since the main focus was on the selection method, we purposely chose a simple replacement method to showcase it. Indeed, future researchers will apply \textit{BinarySelect} with their replacement algorithms for stronger attack research. 

2. \textbf{Human Validation of Choices} - In the pilot study, we compare \textit{BinarySelect} to GreedySelect. While \textit{BinarySelect} clearly exhibits a stronger performance in terms of queries, it is not known to what extent the top or (top X) word is retrieved. Part of this issues lies with the goal of the selection methods. Since selection methods are basing their decision off of classifier feedback rather than human feedback, we cannot simply ask humans which words are most beneficial to the classification. This is because classifiers do not always choose the terms humans consider. This explainabilty of models is an open problem in and of itself. Still incorporation of \textit{BinarySelect} into other downstream related tasks could help verify its selection strength. 
\section{Ethical Considerations}
One must always take into consideration the negative uses of research. This is especially relevant when dealing with adversarial attacks. A malicious user may take \textit{BinarySelect} and use it to improve a system which targets or harasses others. However, we believe the positive uses of our proposed algorithm outweigh the negative uses. This is especially true since this algorithm and code is made available to the public, which allows other researchers to build on or research defenses against it. Furthermore, the second stage of the attack is simple and therefore already known in this space. We believe these reasons along with the potential positives of allowing researchers with low access to computational power, to justify publishing.

\bibliography{main}
\bibliographystyle{acl_natbib}

\newpage
\appendix

\onecolumn
\newpage
\twocolumn

\section{Binary Select}\label{sect:bsalg}
The full algorithm for BinarySelect can be found in Algorithm \ref{alg:bs}.

\begin{algorithm}
\caption{Binary Select}\label{alg:bs}
\begin{algorithmic}[1]
\REQUIRE $text$
\ENSURE $most\_influential\_pos$
%\STATE $most\_influential\_pos \gets 0$
\STATE $Score_{Orig} \gets Classifier(text)$
\STATE $start \gets 0, end  \gets \text{len}(text) - 1$ 
\WHILE{$start \neq end$}
    \STATE $mid \gets (start + end) // 2$
    \STATE $left\_text \gets text[0:mid+1]$
    \STATE $right\_text \gets text[mid+1:]$
    \STATE $Score_{Left} \gets Classifier(left\_text)$
    \STATE $Score_{Right} \gets Classifier(right\_text)$
    \STATE $DropLeft \gets Score_{Orig} - Score_{Left}$
    \STATE $DropRight \gets Score_{Orig} - Score_{Right}$
    \IF{$DropLeft > DropRight$}
        \STATE $end \gets mid$
    \ELSE
        \STATE $start \gets mid + 1$
    \ENDIF
\ENDWHILE
\STATE $most\_influential\_pos \gets start$
%\RETURN $most\_influential\_pos$
\end{algorithmic}
\end{algorithm}
\newpage
\onecolumn
\twocolumn
\section{Related Work}\label{sect:relatedwork}
Adversarial text attacks are useful for testing robustness of models and even in areas of privacy concerns and censorship \cite{xie-hong-2022-differentially}. Adversarial attacks are executed at different levels: 1. Character, 2. Word, 3. Phrase, 4. Sentence, 5. Multi-level. 

Character-level attacks change individual characters in words to cause tokens to become unknown to the target NLP models. These attacks include addition/removal of whitespace \cite{Grndahl2018AllYN}, replacement of visually similar characters \cite{eger-etal-2019-text}, and shuffling of characters \cite{Li2019Textbugger}. Word-level attacks replace words with synonyms that are less known to the target NLP models. The attacks have leveraged Word Embeddings \cite{hsieh-etal-2019-robustness}, WordNet \cite{ren-etal-2019-generating}, and Mask Language Models \cite{li-etal-2020-bert-attack} to find relevant synonyms for replacement. Phrase-level attacks replace multiple consecutive words at once \cite{deng-etal-2022-valcat,lei-etal-2022-phrase}. Sentence-level attacks leverage generation methods to rewrite text in a format that the target NLP model is unfamiliar with \cite{ribeiro-etal-2018-semantically,zhao2018generating}. Multi-level attacks use a combination of the above attacks to cause model failure \cite{formento-etal-2023-using}. We test our proposed methodology at the word-level, however, it could be extended to the character or phrase level easily. 

Adversarial attacks have different levels of knowledge of their target model. White-box attacks are able to leverage complete model information, including the weights of the trained model and architecture \cite{sadrizadeh-2022-block, wang-etal-2022-semattack}. Black-box attacks only have access to a models' confidence level (e.g. probabilities or logits) as well as their output \cite{le-etal-2022-perturbations,jin2020bert}. In the case of text classification that output is the predicted label. Since white-box attacks have access to the weights of a model, they are able to find words to replace or modify very quickly. Black-box attacks however, need many queries since they only have access to classifier confidence which they check when making changes. As noted, our research aims to improve on previous black-box attacks by decreasing the number of queries needed to find the best words to replace.

\section{BS Structure Algorithm}\label{sect:bsstructalg}

Algorithm \ref{alg:bsnode} shows the updated \textit{BinarySelect} algorithm with use of the binary tree (BSNode) structure described in Section \ref{sect:adversarialattack}.

\begin{algorithm}
\caption{Binary Select}\label{alg:bsnode}
\begin{algorithmic}[0]
\REQUIRE text
\ENSURE most\_influential\_pos

\STATE Initialize BS structure with the root node representing the entire text and its corresponding classifier score.
\STATE Initialize most\_influential\_pos to None.
\STATE $ScoreOrig \leftarrow Classifier(text)$
\STATE $DropMax \leftarrow 0$

\WHILE{BS structure is not fully explored}
    \STATE $cur\_node \leftarrow$ BS node with the lowest unexplored probability score
    \IF{cur\_node is a leaf node}
        \STATE Mark cur\_node as explored
        \IF{most\_influential\_pos is None or cur\_node.prob < BS node at most\_influential\_pos.prob}
            \FOR{each word w in cur\_node.data}
                \STATE $Scorew \leftarrow Classifier(text/w)$, where w is the word represented by cur\_node
                \STATE $Dropw \leftarrow ScoreOrig - Scorew$
                \IF{Dropw > DropMax}
                    \STATE $DropMax \leftarrow Dropw$
                    \STATE most\_influential\_pos $\leftarrow$ position of cur\_node in the original text
                \ENDIF
            \ENDFOR
        \ENDIF
    \ELSE
        \STATE Split cur\_node's text segment into two parts
        \STATE Create left and right child nodes in the BS structure for the two parts
        \STATE Mark cur\_node as explored
    \ENDIF
\ENDWHILE

\RETURN most\_influential\_pos
\end{algorithmic}
\end{algorithm}
\section{Experimental Details: Datasets and Classifiers}\label{sect:experimentaldetails}
To verify BinarySelect in an attack setting, we test it and GreedySelect against the following datasets and classifiers.

\subsection{Datasets:} We test the attack on the following datasets, examined in previous attack research \cite{jin2020bert, li-etal-2020-bert-attack}, randomly sampling 1000 examples from each test set:

1. Yelp Polarity - binary sentiment classification, containing texts from Yelp reviews. The labels are positive or negative. The average text lengths are 157 tokens.

2. IMDB - binary sentiment classification, containing text reviews for movies. Labels are positive or negative. The average text lengths are 215 tokens.

3. AG News - A multi-class (Sports, World, Business, Sci/Tech)  dataset containing news texts. The average text lengths are 43 tokens.

\subsection{Classifiers:} We test against 5 classifiers for each dataset, by leveraging pretrained TextAttack \cite{morris2020textattack} and other Huggingface models\footnote{Full list in Appendix \ref{sect:modellist}}:

1. Albert \cite{lan2019albert} - a fine-tuned version of Albert, which shares weights across layers in order to obtain a smaller-memory footprint than BERT.

2. Distilbert \cite{sanh2020distilbert} - a fine-tuned Distilbert model. Distilbert was pretrained using BERT as a teacher for self-supervision and thus is a lighter, faster model than BERT. 

3. BERT \cite{devlin-etal-2019-bert} - a fine-tuned version of BERT-base-uncased. BERT pre-trains on next sentence prediction and masked language modelling tasks to gain an inherent understanding of text.  

%4. XLNet \cite{yang2019xlnet} - a fine-tuned version of XLNet. XLNet achieves higher classification results than BERT in certain tasks by addressing shortcomining in BERT's learning. 

4. RoBERTa \cite{liu2019roberta} - a fine-tuned version of RoBERTa. RoBERTa outperforms BERT in classification tasks, due to different choices in pretraining.  

5. LSTM - LSTM trained on the respective datasets. The trained models are available from TextAttack\footnote{https://textattack.readthedocs.io/en/latest/3recipes/models.html}.

\section{List of Huggingface Models}\label{sect:modellist}
Table \ref{tab:modellist} contains the locations of the different models tested for our attack. 

\begin{table*}[]
    \centering
    \begin{tabular}{c|c|p{7.5cm}}
         & Model & Huggingface Location \\\hline
         \parbox[t]{2mm}{\multirow{5}{*}{\rotatebox[origin=c]{90}{Yelp }}} & Albert & textattack/albert-base-v2-yelp \\
         & Distilbert & randellcotta/distilbert-base-uncased-finetuned-yelp-polarity \\
         & BERT & textattack/bert-base-uncased-yelp \\
         & Roberta & VictorSanh/roberta-base-finetuned-yelp-polarity\\
         & LSTM & lstm-yelp (TextAttack)\\\hline
         \parbox[t]{2mm}{\multirow{5}{*}{\rotatebox[origin=c]{90}{IMDB }}} & Albert & textattack/albert-base-v2-imdb \\
         & Distilbert & textattack/distilbert-base-uncased-imdb \\
         & BERT & textattack/bert-base-uncased-imdb \\
         & Roberta & textattack/roberta-base-imdb \\
         & LSTM & lstm-imdb (TextAttack)\\\hline
         \parbox[t]{2mm}{\multirow{5}{*}{\rotatebox[origin=c]{90}{AG News }}} & Albert & textattack/albert-base-v2-ag-news \\
         & Distilbert & textattack/distilbert-base-uncased-ag-news \\
         & BERT & textattack/bert-base-uncased-ag-news \\
         & Roberta & textattack/roberta-base-ag-news \\
         & LSTM & lstm-ag-news (TextAttack)\\\hline
         
    \end{tabular}
    \caption{The locations of pretrained models tested in our attack research.}
    \label{tab:modellist}
\end{table*}
\begin{table*}[]
    \centering
    \footnotesize
    \begin{tabular}{c|c||cc||cc||cc||cc||cc||}
          &  & \multicolumn{2}{|c||}{Albert} & \multicolumn{2}{|c||}{Distilbert} & \multicolumn{2}{|c||}{BERT} &  \multicolumn{2}{|c||}{Roberta} & \multicolumn{2}{|c||}{LSTM} \\\hline
          &  & GS & BS & GS & BS & GS & BS & GS & BS & GS & BS
         \\\hline

         % yelp 
          \parbox[t]{2mm}{\multirow{4}{*}{\rotatebox[origin=c]{90}{Yelp }}}& Original Acc. & \multicolumn{2}{|c||}{99.8} & \multicolumn{2}{|c||}{95.2} & \multicolumn{2}{|c||}{99.5} &  \multicolumn{2}{|c||}{98.3} & \multicolumn{2}{|c||}{94.7} \\ 
         & Attack Acc. & 71.7 & 76.1 & 63.4 & 73.6 & 76.2 & 80.1 & 80.0 & 85.3 & 44.8 & 64.3\\\cline{2-12}
         & Avg. Queries & 170 & 74 & 172 & 75 & 171 & 74 & 178 & 81 & 166 & 70\\
         & Avg. Q's (Success) & 103 & 44 & 117 & 47 & 107 & 48 & 107 & 52 & 145 & 51\\\hline\hline

        % imdb 
          \parbox[t]{2mm}{\multirow{4}{*}{\rotatebox[origin=c]{90}{IMDB}}}& Original Acc. & \multicolumn{2}{|c||}{97.7} & \multicolumn{2}{|c||}{96.8} & \multicolumn{2}{|c||}{97.9} &  \multicolumn{2}{|c||}{97.6} & \multicolumn{2}{|c||}{84.8}\\ 
         & Attack Acc. & 73.7 & 85.2 & 65.7 & 80.8 & 76.5 & 87.3 & 82.3 & 90.0 & 51.5 & 76.6 \\\cline{2-12}
         & Avg. Queries & 267 & 81 & 266 & 78 & 265 & 82 & 273 & 84 & 254 & 71\\
         & Avg. Q's (Success) & 242 & 51 & 231 & 51 & 254 & 55 & 229 & 57 & 245 & 48\\\hline\hline

        % ag news 
          \parbox[t]{2mm}{\multirow{4}{*}{\rotatebox[origin=c]{90}{AG News}}}& Original Acc. & \multicolumn{2}{|c||}{98.8} & \multicolumn{2}{|c||}{97.4} & \multicolumn{2}{|c||}{99.6} &  \multicolumn{2}{|c||}{99.2} & \multicolumn{2}{|c||}{93.1} \\ 
         & Attack Acc. & 76.6 & 77.6 & 85.5 & 87.6 & 85.7 & 88.1 & 81.9 & 85.1 & 76.7 & 82.1\\\cline{2-12}
         & Avg. Queries & 66 & 53 & 65 & 54 & 69 & 56 & 66 & 56 & 63 & 55 \\
         & Avg. Q's (Success) & 53 & 37 & 53 & 33 & 53 & 38 & 53 & 34 & 54 & 37\\\hline\hline
         
    \end{tabular}
    \caption{Adversarial Attack Results when $k=5$. ``Original Acc.'' is the original accuracy of the model, ``Attack Acc.'' is the model accuracy on the text modified by the attack. ``Avg. Queries'' is the average number of queries used, ``Avg. Q's (Success)'' are the number of queries used for successful attacks. GS - GreedySelect, BS - \textit{BinarySelect}.}
    \label{tab:k5}
\end{table*}

\begin{table*}[]
    \centering
    \footnotesize
    \begin{tabular}{c|c||cc||cc||cc||cc||cc||}
          &  & \multicolumn{2}{|c||}{Albert} & \multicolumn{2}{|c||}{Distilbert} & \multicolumn{2}{|c||}{BERT} &  \multicolumn{2}{|c||}{Roberta} & \multicolumn{2}{|c||}{LSTM} \\\hline
          &  & GS & BS & GS & BS & GS & BS & GS & BS & GS & BS
         \\\hline

         % yelp 
          \parbox[t]{2mm}{\multirow{4}{*}{\rotatebox[origin=c]{90}{Yelp }}}& Original Acc. & \multicolumn{2}{|c||}{99.8} & \multicolumn{2}{|c||}{95.2} & \multicolumn{2}{|c||}{99.5} &  \multicolumn{2}{|c||}{98.3} & \multicolumn{2}{|c||}{94.7}\\ 
         & Attack Acc. & 25.8 & 33.8.0 & 17.5 & 28.3 & 28.5 & 33.6 & 37 & 47.4 & 6.8 & 22.5 \\\cline{2-12}
         & Avg. Queries & 261 & 220 & 233 & 195 & 270 & 220 & 298 & 263 & 184 & 155 \\
         & Avg. Q's (Success) & 197 & 144 & 195 & 138 & 202 & 153 & 210 & 166 & 179 & 112\\\hline\hline

        % imdb 
          \parbox[t]{2mm}{\multirow{4}{*}{\rotatebox[origin=c]{90}{IMDB}}}& Original Acc. & \multicolumn{2}{|c||}{97.7} & \multicolumn{2}{|c||}{96.8} & \multicolumn{2}{|c||}{97.9}  & \multicolumn{2}{|c||}{97.6} &  \multicolumn{2}{|c||}{84.8}\\ 
         & Attack Acc. & 34.2 & 51.4 & 20.9 & 39.5 & 38.5 & 53.7 & 32.7 & 56.4 & 20.2 & 43.3 \\\cline{2-12}
         & Avg. Queries & 369 & 266 & 337 & 231 & 373 & 268 &  384 & 285 & 284 & 196 \\
         & Avg. Q's (Success) & 307 & 167 & 298 & 154 & 299 & 168 &  323 & 180 & 267 & 124\\\hline\hline

           % ag news 
          \parbox[t]{2mm}{\multirow{4}{*}{\rotatebox[origin=c]{90}{AG News}}}& Original Acc. & \multicolumn{2}{|c||}{98.8} & \multicolumn{2}{|c||}{97.4} & \multicolumn{2}{|c||}{99.6} & \multicolumn{2}{|c||}{99.2} & \multicolumn{2}{|c||}{93.1}\\ 
         & Attack Acc. & 23.9 & 23.9 & 33.0 & 32.1 & 32.9 & 34.7 & 28.8 & 29.2 & 21.9 & 22.1 \\\cline{2-12}
         & Avg. Queries & 148 & 155 & 168 & 182 & 176 & 192 & 164 & 175 & 134 & 150 \\
         & Avg. Q's (Success) & 117 & 119 & 135 & 143 & 139 & 146 & 128 & 135 & 112 & 125\\\hline\hline

    \end{tabular}
    \caption{Adversarial Attack Results when $k=30$. ``Original Acc.'' is the original accuracy of the model, ``Attack Acc.'' is the model accuracy on the text modified by the attack. ``Avg. Queries'' is the average number of queries used, ``Avg. Q's (Success)'' are the number of queries used for successful attacks. GS - GreedySelect, BS - \textit{BinarySelect}.}
    \label{tab:k30}
\end{table*}

\begin{table*}[]
    \centering
    \footnotesize
    \begin{tabular}{c|c||cc||cc||cc||cc||cc||}
          &  & \multicolumn{2}{|c||}{Albert} & \multicolumn{2}{|c||}{Distilbert} & \multicolumn{2}{|c||}{BERT} & \multicolumn{2}{|c||}{XLNet} & \multicolumn{2}{|c||}{Roberta} \\\hline
          &  & GS & BS & GS & BS & GS & BS & GS & BS & GS & BS
         \\\hline

         % yelp 
          \parbox[t]{2mm}{\multirow{4}{*}{\rotatebox[origin=c]{90}{Yelp }}}& Original Acc. & \multicolumn{2}{|c||}{99.8} & \multicolumn{2}{|c||}{95.2} & \multicolumn{2}{|c||}{99.5}  & \multicolumn{2}{|c||}{98.3} & \multicolumn{2}{|c||}{94.7}\\ 
         & Attack Acc. & 16.3 & 22.1 & 10.9 & 16.1 & 16.0 & 21.5 & 24.9 & 33.5 & 6.2 & 20.0\\\cline{2-12}
         & Avg. Queries & 295 & 270 & 248 & 233 & 307 & 275 & 348 & 347 & 186 & 194\\
         & Avg. Q's (Success) & 232 & 194 & 219 & 186 & 250 & 199 & 262 & 231 & 181 & 122\\\hline\hline

        % imdb 
          \parbox[t]{2mm}{\multirow{4}{*}{\rotatebox[origin=c]{90}{IMDB}}}& Original Acc. & \multicolumn{2}{|c||}{97.7} & \multicolumn{2}{|c||}{96.8} & \multicolumn{2}{|c||}{97.9} & \multicolumn{2}{|c||}{97.6} & \multicolumn{2}{|c||}{84.8} \\ 
         & Attack Acc. & 24.4 & 37.7 & 12.0 & 27.2 & 24.9 & 37.7 &  17.7 & 42.2 & 19.2 & 38.6 \\\cline{2-12}
         & Avg. Queries & 417 & 356 & 359 & 294 & 425 & 361 &  422 & 383 & 293 & 259 \\
         & Avg. Q's (Success) & 334 & 230 & 326 & 204 & 346 & 242 &  371 & 252 & 270 & 149\\\hline\hline

           % ag news 
          \parbox[t]{2mm}{\multirow{4}{*}{\rotatebox[origin=c]{90}{AG News}}}& Original Acc. & \multicolumn{2}{|c||}{98.8} & \multicolumn{2}{|c||}{97.4} & \multicolumn{2}{|c||}{99.6}  & \multicolumn{2}{|c||}{99.2} & \multicolumn{2}{|c||}{93.1}\\ 
         & Attack Acc. &  17.2 & 18.3 & 23.3 & 23.7 & 23.5 & 24.2 & 18.5 & 20.0 &  16.2 & 16\\\cline{2-12}
         & Avg. Queries & 159 & 169 & 185 & 202 & 195 & 212 & 179 & 192 &  143 & 160\\
         & Avg. Q's (Success) & 132 & 134 & 155 & 164 & 159 & 173 & 152 & 158 & 124 & 139\\\hline\hline
         
    \end{tabular}
    \caption{Adversarial Attack Results when $k=50$. ``Original Acc.'' is the original accuracy of the model, ``Attack Acc.'' is the model accuracy on the text modified by the attack. ``Avg. Queries'' is the average number of queries used, ``Avg. Q's (Success)'' are the number of queries used for successful attacks. GS - GreedySelect, BS - \textit{BinarySelect}.}
    \label{tab:k50}
\end{table*}

\begin{table*}[]
    \centering
    \footnotesize
    \begin{tabular}{c|c||cc||cc||cc||cc||cc||}
          &  & \multicolumn{2}{|c||}{Albert} & \multicolumn{2}{|c||}{Distilbert} & \multicolumn{2}{|c||}{BERT} &  \multicolumn{2}{|c||}{Roberta} & \multicolumn{2}{|c||}{LSTM}\\\hline
          &  & GS & BS & GS & BS & GS & BS & GS & BS & GS & BS
         \\\hline

         % yelp 
          \parbox[t]{2mm}{\multirow{4}{*}{\rotatebox[origin=c]{90}{Yelp }}}& Original Acc. & \multicolumn{2}{|c||}{99.8} & \multicolumn{2}{|c||}{95.2} & \multicolumn{2}{|c||}{99.5} & \multicolumn{2}{|c||}{98.3} & \multicolumn{2}{|c||}{94.7} \\ 
         & Attack Acc. & 5.3 & 5.2 & 5.1 & 5.2 & 4.6 & 5.2 & 9.2 & 8.0 & 5.3 & 5.3\\\cline{2-12}
         & Avg. Queries & 372 & 427 & 271 & 313 & 372 & 415  & 476 & 592 & 196 & 421 \\
         & Avg. Q's (Success) & 336 & 380 & 268 & 310 & 339 & 378 & 420 & 548 & 196 & 421\\\hline\hline

        % imdb 
          \parbox[t]{2mm}{\multirow{4}{*}{\rotatebox[origin=c]{90}{IMDB}}}& Original Acc. & \multicolumn{2}{|c||}{97.7} & \multicolumn{2}{|c||}{96.8} & \multicolumn{2}{|c||}{97.9} & \multicolumn{2}{|c||}{97.6} & \multicolumn{2}{|c||}{84.8}\\ 
         & Attack Acc. & 4.7 & 4.7 & 3.4 & 3.8 & 3.8 & 4.8 &  2.7 & 3.1 & 15.2 & 15.2 \\\cline{2-12}
         & Avg. Queries & 571 & 709 & 407 & 526 & 578 & 746 & 489 & 724 & 353 & 756\\
         & Avg. Q's (Success) & 549 & 670 & 405 & 520 & 550 & 705 & 486 & 713 & 353 & 756\\\hline\hline

             % ag news 
          \parbox[t]{2mm}{\multirow{4}{*}{\rotatebox[origin=c]{90}{AG News}}}& Original Acc. & \multicolumn{2}{|c||}{98.8} & \multicolumn{2}{|c||}{97.4} & \multicolumn{2}{|c||}{99.6}  & \multicolumn{2}{|c||}{99.2} & \multicolumn{2}{|c||}{93.1} \\ 
         & Attack Acc. & 17.1 & 18.0 & 23.2 & 23.2 & 23.1 & 24.2 & 18.3 & 20.4 & 16.2 & 16.0\\\cline{2-12}
         & Avg. Queries & 159 & 169 & 186 & 202 & 196 & 214 & 180 & 193 & 143 & 161 \\
         & Avg. Q's (Success) & 133 & 134 & 156 & 164 & 161 & 172 & 153 & 157 & 124 & 139\\\hline\hline
         
    \end{tabular}
    \caption{Adversarial Attack Results when $k=\text{ALL}$. ``Original Acc.'' is the original accuracy of the model, ``Attack Acc.'' is the model accuracy on the text modified by the attack. ``Avg. Queries'' is the average number of queries used, ``Avg. Q's (Success)'' are the number of queries used for successful attacks. GS - GreedySelect, BS - \textit{BinarySelect}.}
    \label{tab:kall}
\end{table*}

\section{$k$ Results}\label{app:kresults}
We generate similar tables to Table \ref{tab:attackResults} for $k = \{5, 15, 30, 50, \text{ALL}\}$. 
Table \ref{tab:k5} is $k=5$, Table \ref{tab:attackResults} is $k=15$, Table \ref{tab:k30} is $k=30$, Table \ref{tab:k50} is $k=50$, and Table \ref{tab:kall} is $k=\text{ALL}$.

%\section{\textit{BinarySelect} with Binary Tree Structure}

\section{EDR Charts}
\begin{figure}
    \centering
    \begin{subfigure}[b]{0.95\columnwidth}
        \centering
        \fbox{\includegraphics[width=\textwidth]{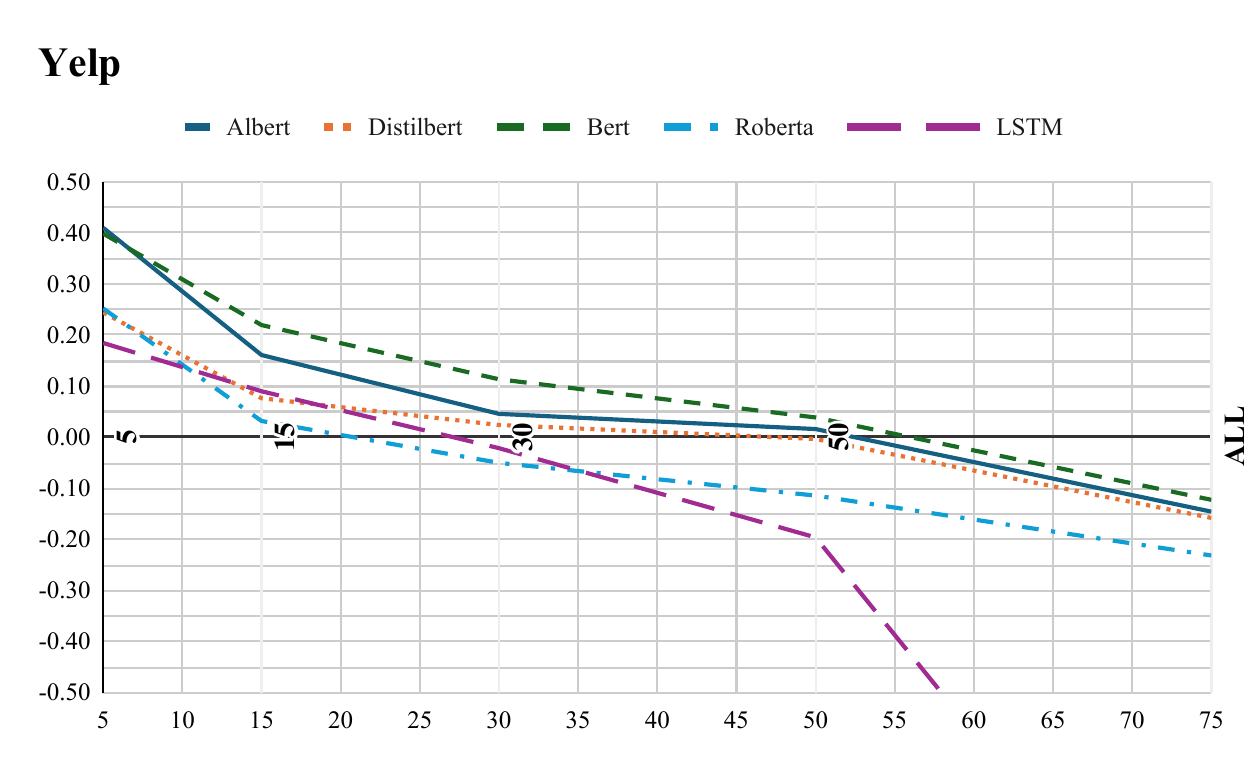}}
    \end{subfigure}
    \begin{subfigure}[b]{0.95\columnwidth}
        \centering
        \fbox{\includegraphics[width=\textwidth]{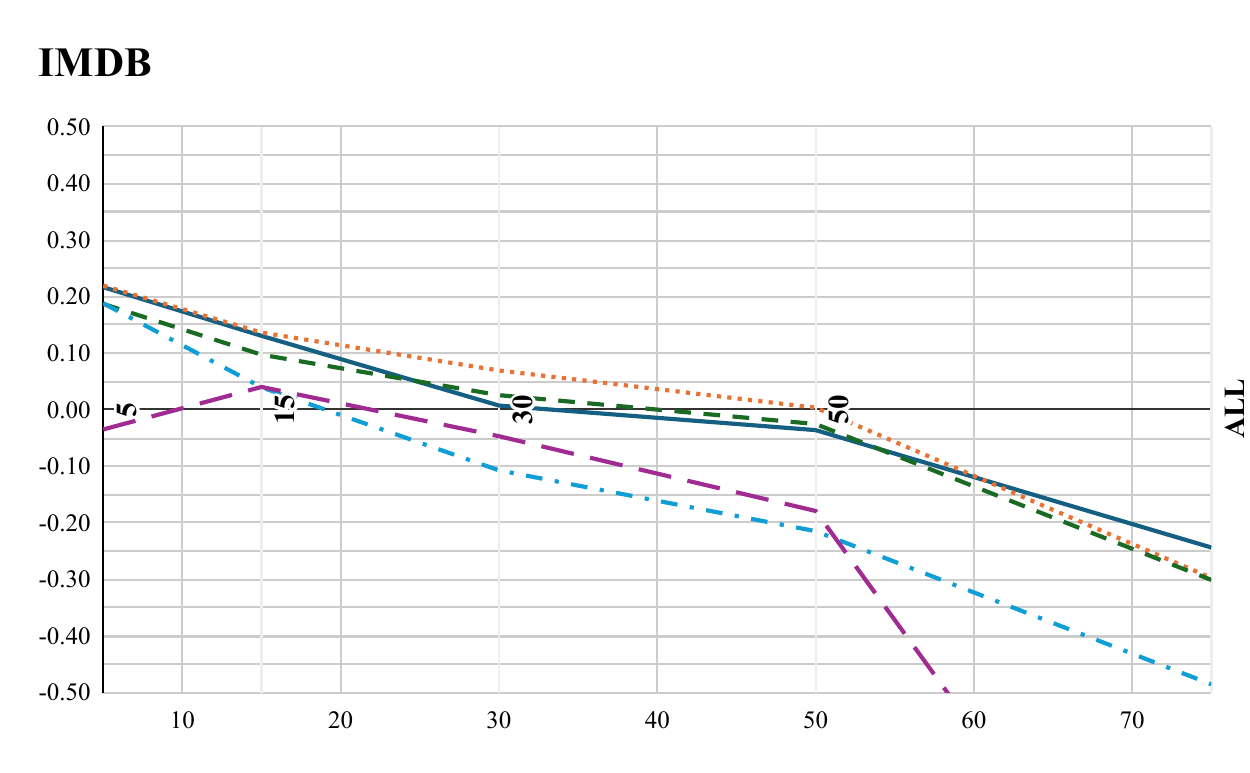}}
    \end{subfigure}
    \begin{subfigure}[b]{0.95\columnwidth}
        \centering
        \fbox{\includegraphics[width=\textwidth]{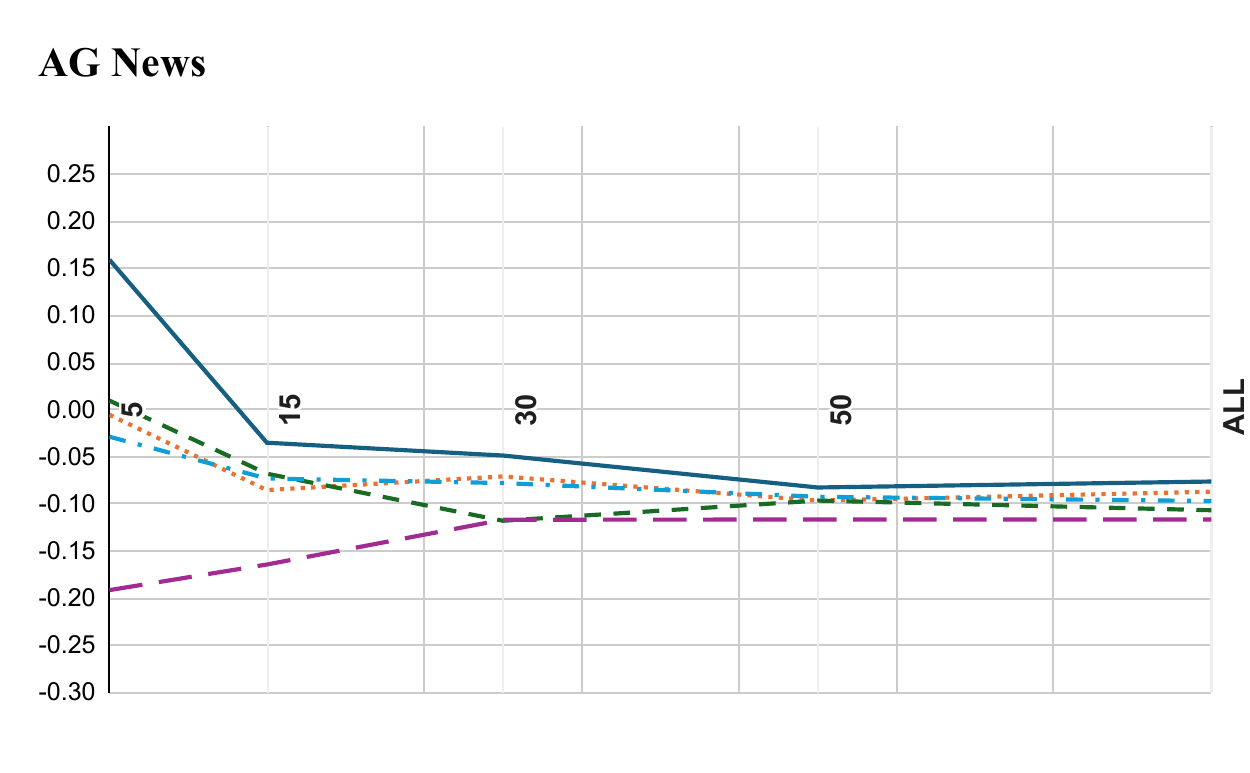}}
    \end{subfigure}
    \caption{Effect of $k$ Values on EDR (Equation \ref{eq:edr}) for the all attacks. Positive values indicate a better trade-off between reduction in queries versus loss of accuracy drop for BS.}
    \label{fig:kchartsall}
\end{figure}

Figure \ref{fig:kchartsall} shows the EDR values for different $k$ values for both success and failed attacks. Trends are similar to Figure \ref{fig:kcharts}, although the failed attacks cause a lesser trade off for larger $k$.

\end{document}